\input amssym.def 
\input amssym 
\newcommand{\Bbc}[1]{{\Bbb{#1}}} 
\newcommand{\Beq}{\begin{equation}}
\newcommand{\Eeq}{\end{equation}}
\newcommand{\Beqa}{\begin{eqnarray}}
\newcommand{\Eeqa}{\end{eqnarray}}
\newcommand{\End}{\nonumber\\}

\newcommand{\Lbb}{\Bigl(}

\newcommand{\Rbb}{\Bigr)}

\newcommand{\Lsa}{\bigl[}
\newcommand{\Rsa}{\bigr]}
\newcommand{\Lsb}{\Bigl[}
\newcommand{\Rsb}{\Bigr]}

\newcommand{\Seto}{\{}
\newcommand{\Setc}{\}}
\newcommand{\Real}{\Bbc{R}}
\newcommand{\Comp}{\Bbc{C}}
\newcommand{\Half}{{\textstyle\frac{1}{2}}}
\newcommand{\Frac}[2]{{\textstyle\frac{#1}{#2}}}
\newcommand{\Ft}[1]{\hat{#1}}
\newcommand{\Bare}{{\Bbb{R}}_S}

\newcommand{\Rmn}[2]{\Bare^{#1,#2}}

\newcommand{\Row}[2]{#1^1,\dots,#1^{#2}}
\newcommand{\Rowrow}[4]{#1^1,\dots,#1^{#2};#3^1,\dots,#3^{#4}}
\newcommand{\Summun}{\sum_{\mu \in M_n}}
\newcommand{\Part}[1]{\frac{\partial}{\partial #1}}
\newcommand{\Pdvar}[2]{\frac{\partial #1}{\partial #2}}
\newcommand{\CalO}{{\cal O}}
\newcommand{\Bint}{\int_{{\cal B}}}
\newcommand{\Ebold}{{\Bbc E}}
\newcommand{\Refeq}[1]{equation~(\ref{#1})}

\newcommand{\xthr}{(\Rowrow{x}m{\theta}n)}

\newcommand{\One}{1\!\!1}

\title{Stochastic Calculus and Anticommuting Variables}
\author{Alice Rogers\thanks{Research supported by a Royal Society
University Research Fellowship\hfill}\\
Department of Mathematics\\
King's College\\
Strand\\
London WC2R 2LS}
\date{27th September 1994\\ \ \\kcl-th-94-16, hep-th/9409162\\\ \
\\Invited lecture given at meeting on `Espaces de Lacets', Institut de
Recherche Math\'ematique Advanc\'ee, Universit\'e Louis Pasteur,
Strasbourg, June 1994}
\documentstyle[12pt]{article}
\begin{document}
\bibliographystyle{plain}
\maketitle
\begin{abstract}
A theory of integration for anticommuting paths is described. This is
combined with standard It\^o calculus to give a geometric theory of
Brownian paths on curved supermanifolds.
\end{abstract}
This lecture concerns a generalisation of Brownian motion and It\^o
calculus to include paths in spaces of anticommuting variables. The
motivation for this work comes originally from physics, where
anticommuting variables were first introduced by Martin \cite{Martin} in
order to extend Feynman's path integral methods to Fermionic systems.
Subsequently various geometric applications to this approach have
emerged, and additionally anticommutiong variables have been found to
play a significant r\^ole in the quantizaton of systems with gauge
symmetry.
\section{Functions of anticommuting variables}
Suppose that $\Row{\theta}n$ are $n$ anticommuting variables, so that
\Beq
\theta^i \theta^j = - \theta^j \theta^i.
\Eeq
Then, since $(\theta^i)^2$ is zero, the natural space of functions to
consider is the $2^n$-dimensional space of functions of the form
\Beq
f(\Row{\theta}n) = \Summun  f_{\mu} \theta^{\mu}
\Eeq
where $\mu$ denotes a multi-index $\mu_1\dots \mu_{|\mu|}$, with $1\leq
\mu_1 < \dots < \mu_{|\mu|} \leq n$, $M_n$ denotes the set of all such
multi-indices (including the empty multi-index $\emptyset$)
and $\theta^{\mu}
= \One\theta^{\mu_1}\dots\theta^{\mu_{|\mu|}}$. The detailed nature of
this function space is determined by the choice of space in which the
coefficient functions $f_{\mu}$ lie. This may be the real numbers, the
complex
numbers or some space of commuting and anticommuting variables.

The importance of such functions for Fermionic physics is that they
allow realisations of the canonical anticommutation relations
\Beq
\psi^i\psi^j + \psi^j\psi^i = 2\delta^{ij}
\Eeq
for Fermionic operators, for instance by setting
\Beq
\psi^i = \theta^i + \Pdvar{}{\theta^i}.
\label{PSIeq}\Eeq
The partial derivative in this expression is defined by setting
\Beqa
\Frac{\partial \theta^{\mu}}{\partial \theta^j} &=& (-1)^{\ell-
1}\theta^{\mu_1}
\dots
\theta^{\mu_{\ell-1}}\theta^{\mu_{\ell+1}} \dots \theta^{\mu_{|\mu|}}
{\rm\ if\ }j = \mu_{\ell}{\rm\ for\ some\ } \ell, 1 \leq \ell \leq
|\mu|,
\End
\Frac{\partial \theta^{\mu}}{\partial \theta^{j}} &=& 0 {\rm\ \
otherwise}.
\Eeqa
Following Berezin \cite{Berezin1}, integration is defined by setting
\Beq
\int d^n\theta\, f(\Row{\theta}n) = f_{1\dots n}
\Eeq
if
\Beq
f(\Row{\theta}n) = f_{1\dots n} \theta^1\dots\theta^n +
\mbox{ lower order terms}.
\Eeq
This integral is formal, but has many useful properties. For instance,
suppose that the Fourier transform $\Ft{f}$ of the function $f$ is
defined by
\Beq
\Ft{f}(\rho) = \Bint d^n\theta \, \exp (-i\rho.\theta) f(\theta),
\Eeq
(with $\rho.\theta=\sum_{j=1}^n \rho^j\theta^j$); then a direct analogue
of the Fourier inversion formula can be established by explicit
calculation. (Here, and in the following, it will be assumed for
simplicity that $n$ is even, because in this case many factors of $i$
can be avoided in the formulae.)

If $\CalO$ is
a linear operator on the space of functions defined above, its kernel is
the function $\CalO(\Row{\theta}n,\Row{\phi}n)$ of $2n$ anticommuting
variables which satisfies
\Beq
\CalO f(\Row{\theta}n) = \int d^n\phi\, \CalO(\Row{\theta}n,\Row{\phi}n)
f(\Row{\phi}n).
\Eeq
A particular example of such a kernel is the delta function (that is,
the kernel of the identity operator) which takes the form
\Beqa
\delta(\theta-\phi) &=& \Bint d^n \rho \exp (-i\rho.(\theta-\phi))  \End
  &=& \Pi_{j=1}^n(\theta^j - \phi^j).
\Eeqa
In most application both anticommuting and commuting variables are
required, so that one works on what has become known as a superspace.
The notation $\Rmn mn$ will be used for a superspace with $m$ commuting
coordinates and $n$ anticommuting coordinates, with a typical point
denoted $\xthr$.
\section{Fermionic Brownian Motion}
Fermionic Brownian motion is obtained by defining a measure (in a
generalised sense) on the space of paths in $\Rmn 0n$. First, an
abstract ``Grassmann measure'' on the space $(\Rmn 0n)^A$ of functions
$\gamma:A
\to \Rmn 0n$ is defined using the converse of the idea of the Kolmogorov
construction, so
that measures are {\it defined} by their finite-dimensional marginal
distributions. The ingredients for such a measure are thus a collection
of functions $\Seto f_J \Setc$ corresponding to the finite subsets $J$
of $A$; the function $f_J$ has domain $(\Rmn 0n)^J$ and satisfies the
condition
\Beq
\int d^n\theta_1 \dots d^n\theta_{|J|} f_J(\theta_1, \dots,
\theta_{|J|}) = 1.
\Eeq
(Here $\theta_1, \dots, \theta_{|J|}$ are $n$-vectors.)
The functions also satisfy a consistency condition: if $J=\Seto
t_1,\dots,t_N \Setc$ and $J' = \Seto t_1\, \dots, t_{N-1} \Setc$, then
\Beq
\int d^n \theta_N f_J(\theta_1, \dots, \theta_N) = f_{J'}(\theta^1,
\dots,  \theta_{N-1}).
\Eeq
A Grassmann random variable is then defined to be any function on
$(\Rmn 0n)^A$ which can be integrated, possibly by some limiting
process. When $J$ is a finite subset of $A$ and $g$ is a function on
$(\Rmn 0n)^J$, then $g$ is certainly a Grassmann random variable, with
expectation defined to be
\Beq
\Ebold(g) = \int d^n\theta_1 \dots d^n\theta_{|J|}
f_J(\theta_1,\dots,\theta_{|J|}) g(\theta_1,\dots,\theta_{|J|}).
\Eeq
Further details of the concept of Grassmann measure, and the related
concept of Grassmann stochastic process may be found in \cite{GBM}.

Fermionic Brownian motion is an example of a Grassmann stochastic
process; it is derived from Grassmann Wiener measure, which is a measure
on
$(\Rmn 0n)^{(0,\infty)}$ with finite distributions
\Beq
f_J(\theta_1,\rho_1,\dots,\theta_N,\rho_N) = \exp -i\Lsa \sum_{r=1}^N
\rho_r.(\theta_r - \theta_{r-1}) \Rsa
\Eeq
when $J=\Seto t_1, \dots, t_N \Setc $ with $0<t_1, \dots, t_N \leq t$
(and
$\theta_0=0$). (The only $t_r$-dependence in the measure is in the time
ordering.) The corresponding $2n$-dimensional stochastic process
$(\theta_t,\rho_t)$ will be referred to as Fermionic Brownian motion.

This measure is essentially composed of $\delta$-functions,
corresponding to the fact that the free Fermionic Hamiltonian is zero.
Although in many ways this measure resembles conventional Wiener measure
(which is based on the heat kernel of the free Bosonic Hamiltonian (or
flat Laplacian) $-\Half \partial_i\partial_i$), it differs from it in
that the Fourier transform or phase space variables $\rho$ are included.
This means that a Feynman-Kac formula for differential operators of all
orders can be developed.  Explicitly, if $H$ is the operator
\Beq
H  = V(\psi) = \Summun V_{\mu}\psi^{\mu}
\Eeq
(where, as in \Refeq{PSIeq}, $\psi^j =  \theta^j +
\Pdvar{}{\theta^j})$, then
\Beq
\exp(-Ht)\, f(\theta) = \Ebold \Lsb \exp-\Lbb \int_0^t V(\omega(s)) ds
\Rbb f(\theta + \theta(t)) \Rsb
\Eeq
where $\omega(t) = \theta(t) + i \rho(t)$, as is proved in \cite{GBM}.
\section{Superspace Brownian motion}
By mutiplying together finite-dimensional marginal distributions,
Fermionic Wiener measure may be combined with conventional Wiener
measure to give a notion of Brownian paths in superspace. These Brownian
paths can then be used to develop Feynman-Kac formulae for a number of
diffusion operators, as will be seen in later sections. The first step
is to
incorporate stochastic calculus into the framework of superspace
Brownian paths.

There is no useful analogue of the It\^o integral along Fermionic
Brownian paths, because they are too irregular. However, since the
motivation for considering these paths is to study diffusions, and the
measure (incorporating the phase space variables $\rho$) is sufficient
to handle differential operators of all orders, this does not matter.
Also, Fermionic Brownian motion can be combined with conventional
stochastic calculus, so that (if $b_t$ denotes Brownian motion on
$\Real^m$) integrals of the form
\Beq
Z_t - Z_0 = \int_0^t A_s \, ds + \int_0^t \sum_{a=1}^m C_{sa}\, db_s^a
\Eeq
can be constructed when $A_s, C_s^{a}$ are suitably regular adapted
stochastic processes on super Wiener space, and $db_s$ are conventional
Brownian increments. (Full details may be found in \cite{SCSONE}.)

Sufficiently well-behaved functions $G(\Row {Z_t}{p})$ of such integrals
satisfy the It\^o formula
\Beqa
&&\Ebold \Lbb G(\Row{Z_t}{p}) \Rbb - G(\Row{Z_0}{p}) \End
&=& \Ebold \Lbb \int_0^t \sum_{j=1}^p A^j_s \partial_j G(\Row{Z_s}p) +
\Half \sum_{a=1}^m \sum_{j=1}^p \sum_{k=1}^p C_{sa}^j
C_{sa}^k \partial_k \partial_j G(\Row{Z_s}p) ds \End
\Eeqa
where the partial derivative $\partial_j$ has the same Grassmann parity
as $Z^j_t$.  This It\^o formula can be used to give Feynman-Kac formulae
for diffusions with drift in the usual way.
\section{Stochastic differential equations for superspace Brownian
paths}\label{SDEsec}
So far, it has been shown how Fermionic Brownian motion allows one to
construct Feynman-Kac formulae for differential operators on functions
on $\Rmn mn$ which are second order in the even variables, with second
order part simply the flat Laplacian $-\Half
\sum_{i=1}^m\partial_i\partial_i$. In order to extend this approach to
study operators
in curved space, as is necessary both for geometrical applications
and for the study of Fermions in a gravitational background, stochastic
differential equations are required. As is shown in \cite{SCSONE},
stochastic differential equations of the form
\Beqa
dZ^j_s &=& \sum_{a=1}^m A_a^{j}(Z_s,\theta_s, \rho_s,s) db^a_s +
B^j(Z_s,\theta_s,\rho_s,s) ds \End
Z_0^j &=& Z^j,\qquad j=1,\dots,p
\Eeqa
have unique solutions provided that the functions $A_a^j$ and $B^j$ are
suitably regular.

The relevance of stochastic differential equations to this article is
that they allow one to extend the scope of these stochastic methods for
studying diffusions; as the following example shows, a wide class of
second-order elliptic operators can be studied. (This example is a
standard example from conventional stochastic calculus \cite{IkeWat},
included for those who are unfamiliar with this technique.)

Suppose that $x^i_t$ satisfies the stochastic differential equation
\Beqa
dx^i_t &=& \sum_{a=1}^m \Lsa V^i_a(x_t)db^a_t + \Half V^j_a(x_t)\,
\partial_j V_a^i(x_t) dt \Rsa \End
x^i_0 &=& x^i \qquad i=1, \dots ,p.
\Eeqa
It will now be shown that
\Beq
\Ebold \Lsa f(x_t) \Rsa = \exp ( -Ht) f(x)
\Eeq
where $H$ is the differential operator
\Beq
H = - \Half \sum_{a=1}^m  V^a V^a ,
\Eeq
with $V_a = \sum_{i=1}^p V^i_a \partial_i$.
This result is proved by applying the It\^o formula  to $f(x_t)$, which
gives
\Beqa
f(x_t) - f(x) &=& \int_0^t \sum_{a=1}^m V^i_a(x_s)\partial_i f(x_s)
db_s^a
+ \int_0^t \Half \sum_{a=1}^m (V^i_a V^j_a \partial_i  \partial_j)f(x_s)
ds \End
&& +\int_0^t \Half \sum_{a=1}^m\sum_{i=1}^p  \partial_j V_a^i V_a^j
\partial_i f(x_s) ds.
\Eeqa
Thus, taking expectations of both sides,
\Beq
\Ebold \Lsa f(x_t) \Rsa - f(x) = \int_0^t \Ebold \Lsa \Half \sum_{a=1}^m
V^aV^af(x_s) \, \Rsa ds
\Eeq
so that, if the operator $U_t$ is defined by
\Beq
U_t(g)(x) = \Ebold [g(x_t)],
\Eeq
then we have shown that
\Beq
U_tf(x) - f(x) = \int_0^t -U_s H(f)(x) ds
\Eeq
from which may be deduced that $U_t = \exp (-Ht)$ as required.

\section{Brownian motion on supermanifolds}
The technique of the previous section gives a Feynman-Kac formula for
the operator $H = \sum_{a=1}^mV^aV^a$. In curved space the Laplacian is
closely related to an operator of this form, and thus, following
Elworthy \cite{Elwort} and Ikeda and Watanabe \cite{IkeWat}, the
solutions to carefully
constructed stochastic differential equations can be used to study the
heat kernels of the various Laplacians which occur on Riemannian
manifolds. This approach can be extended to supermanifolds, as will now
be described, leading to Feynman-Kac formulae for further Hamiltonians.

The appropriate supermanifold (for both geometric and physical
applications) is constructed from a classical Riemannian manifold
$(M,g)$ of dimension $m$ together with a smooth $n$-dimensional complex
vector bundle $E$ over $M$. The supermanifold $S(M,E)$ is of
dimension $(m,m+n)$; it can be specified by stating the transition
functions on overlapping coordinate patches, the actual supermanifold
then being realised by a patching construction \cite{ERICE}. Suppose
that $\Seto (U_{\alpha},\phi_{\alpha}) \Setc $ is an atlas of charts on
$M$ (with each $U_{\alpha}$ also a trivialisation neighbourhood for
$E$) and that $g_{\alpha\beta}:U_{\alpha} \cap U_{\beta} \to Gl(n,
\Comp)$ are the transition functions for the bundle $E$. Then the
supermanifold is defined by taking $m$ even
coordinates $x^i_{\alpha},i=1, \dots m$, $m$ odd coordinates
$\theta_{\alpha}^i, i=1, \dots m$ and $n$ further odd coordinates
$\eta^r_{\alpha}, r=1, \dots, n$, with transition functions
\Beqa
x^i_{\alpha} &=& \phi_{\alpha}^i \circ \phi_{\beta}^{-1}(x_{\beta}) \End
\theta^i_{\alpha} &=& \sum_{j=1}^m\frac{\partial (\phi_{\alpha}^i
\circ \phi_{\beta}^{-1})}{\partial x^j}(x_{\beta}) \theta^j_{\beta} \End
\eta^r_{\alpha} &=& \sum_{s=1}^n g_{\alpha\beta}{}^r{}_s
(\phi_{\beta}^{-1}(x_{\beta})) \eta^s_{\beta}.
\Eeqa
Equipped with this supermanifold it can be seen that $E$-valued forms
on $M$ (which locally can be expresses in the form
\Beq
f(x) = \Summun \sum_{r=1}^{n} f_{\mu\,r}(x) dx^{\mu} e^r
\Eeq
where $(e^r)$ is the local trivialisation basis of $E$)
correspond to smooth functions on the supermanifold $S(M,E)$ which are
linear in the $\eta$, these functions having the local form
\Beq
f(x, \theta,\eta )= \Summun \sum_{r=1}^{n} f_{\mu\,r}(x)
\theta^{\mu}\eta^r.
\Eeq
Under this correspondence the Laplace-Beltrami $L$ operator on twisted
forms becomes simply a differential operator, and so superspace
stochastic calculus can be applied to study the corresponding heat
kerenel.
Specifically \cite{SCSTWO} \eject
\Beqa
L &=& \Half(d + \delta)^2 \End
  &=& -\Half \Lsb \sum_{a=1}^m W_aW_a + \Frac14
[\psi^i,\psi^j]F_{ij\,r}{}^s\, \eta^r \Part{\eta^s} \End
&&+ R^j_i (x) \theta^i \Part{\theta^j} - \Half R_{ki}{}^{j\ell}
\theta^i\theta^k \Part{\theta^j}\Part{\theta^{\ell}} \Rsb \End
\Eeqa
where the summation convention is used and
\Beq
W_a = e^i_a \Part{x^i} - e^j_a e^k_b \Gamma_{jk}^i \Part{e^i_b} - e^j_a
\theta^k \Gamma_{jk}^i \Part{\theta^i} - e^j_a \eta^r A_{j\,r}^s
\Part{\eta^s}.
\Eeq
(Here $A$ and $F$ are respectively the connection and curvature on $E$,
and the
supermanifold $S(M,E)$ is extended to include $O(M)$, the bundle of
orthonormal frames; on $O(M)$ the local coordinates are
$x^i,e^i_a,a,i=1, \dots,m$, corresponding to an
orthonormal basis $(e_a)$ of the tangent space, expanded in the basis of
coordinate derivatives as $e_a=e_a^{i}\Part{x^{i}}$). The space of
functions on which $W_aW_a$ acts is the space of functions on this
extended supermanifold which are independent of the $e^i_a$.

Now suppose that $(x_t^{i},\theta_t^{i}, \rho_t^{i}, \eta_t^{i},
e^{i}_{a\,t})$ are solutions to the stochastic differential equation
corresponding to $W_a$ \cite{SCSTWO}. Then, arguing as in
section~\ref{SDEsec}, it can be shown that
\Beq
\exp(-Lt)g(x,\theta,\eta) = \Ebold \Lsb g(x_t, \theta_t, \eta_t) \Rsb.
\Eeq
\section{Application to the index theorem}
In general the stochastic differential equations of the previous section
cannot be solved in closed form; however they can be applied to give
estimates, particularly as $t$ tends to zero. In addition to the usual
estimate $b_t \sim \surd t$, one has $\theta_t \sim 1$ and $\eta_t \sim
1$.

One example of the application of these methods is the so-called
supersymmetric proof of the Atiyah-Singer index theorem. Using the
formula of McKean and Singer \cite{McKSin}, and considering the case of
the twisted Hirzebruch complex, this amounts to evaluating the
`supertrace' of $\exp(-Lt)$ (that is, the trace of $\tau \exp(-Lt)$
where $\tau$ is an involution on forms) in the limit as $t$ tends to
zero. Using the identification established above of twisted forms with
functions on the supermanifold $S(M,E)$, the required supertrace can
be expressed as an integral of the kernel of $\exp(-Lt)$. In the
original supersymmetric proofs of the index theorem of Alvarez-Gaum\'e
\cite{Alvare} and of Friedan and Windy \cite{FriWin}
somewhat heuristic arguments were used to show that when evaluating this
supertrace the operator $L$ can be replaced by a simpler operator whose
kernel is well known. The stochastic approach described here allows this
step to be made rigorous \cite{SCSTWO}.

Asymptotic expressions for heat kerenels are also important in quantum
gravity \cite{DeWitt2}. The stochastic calculus in this paper should
make it possible
to extend this approach to gravitational theories involving Fermions.
%

\end{document}